\documentclass[aps,twocolumn,pra,superscriptaddress,amsmath,amssymb]{revtex4-1}
\usepackage{dcolumn}
\usepackage{bm}
\usepackage{amsmath}
\usepackage{txfonts}
\usepackage[T1]{fontenc}
\usepackage{xspace}
\usepackage{ulem}
\usepackage{braket}
\ifx\pdfoutput\undefined
\usepackage[dvipdfmx]{graphicx}
\usepackage[dvipdfmx]{hyperref}
\usepackage[dvipdfmx]{color}
\else
\usepackage{graphicx}
\usepackage{hyperref}
\usepackage{color}
\fi
\begin{document}
\let\emph\textit

\title{
  Antiferromagnetically ordered state in the half-filled Hubbard model
  on the Socolar dodecagonal tiling
}

\author{Akihisa Koga }
\affiliation{
  Department of Physics, Tokyo Institute of Technology,
  Meguro, Tokyo 152- 8551, Japan
}

\date{\today}
\begin{abstract}
  We investigate the antiferromagnetically ordered state
  in the half-filled Hubbard model on the Socolar dodecagonal tiling.
  When the interaction is introduced,
  the staggered magnetizations suddenly appear,
  which results from the existence of the macroscopically degenerate states in
  the tightbinding model.
  The increase of the interaction strength monotonically increases
  the magnetizations although its magnitude depends on the local environments.  
  Magnetization profile is discussed in the perpendicular space.
  The similarity and difference are also addressed in magnetic properties
  in the Hubbard model on the Penrose, Ammann-Beenker, and
  Socolar dodecagonal tilings.
\end{abstract}
\maketitle

\section{Introduction}
Quasiperiodic systems have attracted much interest since
the first discovery of the quasicrystal Al-Mn\cite{Shechtman}.
One of the most interesting examples is the Au-Al-Yb alloy
with Tsai-type clusters~\cite{Ishimasa_2011}.
In the quasicrystal Au$_{51}$Al$_{34}$Yb$_{15}$,
quantum critical behavior appears at low temperatures
while heavy fermion behavior appears in the approximant
Au$_{51}$Al$_{35}$Yb$_{14}$~\cite{Deguchi_2012}.
This implies that electron correlations as well as quasiperiodic structures
play an important role in understanding low temperature properties
in the system~\cite{Watanabe,Andrade,Takemori_2015,Takemura_2015,Otsuki,Shinzaki_2016}.
Furthermore, the superconducting state has recently been observed
in the quasicrystal Al-Zn-Mg~\cite{Kamiya_2018},
which stimulates further investigations on
spontaneously symmetry breaking state
in correlated electron systems
on quasiperiodic lattices~\cite{Sakai2017,Sakai2019,Inayoshi_2020}.

Up to now, the magnetically ordered states have not been observed in the quasicrystals,
but have recently been realized in the approximants
$\rm Cd_6Tb$~\cite{Tamura_2010},
Au-Al-Gd~\cite{PhysRevB.93.024416} and Au-Al-Tb~\cite{PhysRevB.98.220403}.
This accelerates the experimental and theoretical investigations
on the magnetic properties in the quasiperiodic lattices.
The Ising~\cite{Okabe,PhysRevB.44.9271,Komura},
Hubbard~\cite{Jagannathan_Schulz_1997,Koga_Tsunetsugu_2017,ABKoga}, and
Heisenberg~\cite{Wessel_2003,Jagannathan_2007} models on the two-dimensional
quasiperiodic systems
have been discussed so far.
In our previous papers, we have considered magnetic properties
in the half-filled Hubbard models on the Penrose~\cite{Koga_Tsunetsugu_2017}
and Ammann-Beenker~\cite{ABKoga} tilings.
Both models have some common magnetic properties.
One of them is that the antiferromagnetically ordered
state is realized without a uniform magnetization in the thermodynamic limit
since each tiling is bipartite and has no sublattice imbalance.
Namely, the magnetization profile in the strong coupling regime
is almost classified into some groups, depending on the coordination number for the site.
Another is an interesting spatial distribution in the staggered magnetization
in the weak coupling limit.
This originates from the existence of the delta-function peak at $E=0$
in the noninteracting density of states.
Nevertheless, a distinct behavior appears in the perpendicular space;
the superlattice structure in the magnetic profile
appears in the Ammann-Beenker tiling case
while it does not appear in the other.
Therefore, by examining the Hubbard model on another bipartite lattice,
it should be instructive to discuss the difference in magnetic properties.

In this study, we focus on the Socolar dodecagonal tiling~\cite{Socolar_1989}
to discuss magnetic properties in the Hubbard model at half filling.
First, we consider the tightbinding Hamiltonian to demonstrate
the existence of the macroscopically degenerate states at $E=0$.
By using the inflation-deflation rule for the tiling,
we examine these confined states to obtain the lower bound of the fraction.
The effects of the Coulomb interactions are studied
by means of the real-space Hartree approximations.
Then we clarify whether or not the superlattice structure appears
in the magnetic profile in the system.

The paper is organized as follows.
In. Sec.~\ref{model},
we introduce the half-filled Hubbard model on the Socolar dodecagonal tiling.
In. Sec.~\ref{conf}, we study the confined states with $E=0$,
which should play an important role for magnetic properties
in the weak coupling limit.
We discuss how the antiferromagnetically ordered state
is realized in the Hubbard model in Sec.~\ref{results}.
The crossover behavior in the ordered state is addressed,
by mapping the spatial distribution of the magnetization
to the perpendicular space.
A summary is given in the last section.

\section{Model and Hamiltonian}\label{model}
We study the Hubbard model on the Socolar dodecagonal
tiling~\cite{Socolar_1989},
which should be given by the following Hamiltonian,
\begin{eqnarray}
  H&=&-t\sum_{(ij),\sigma}\left(c_{i\sigma}^\dag c_{j\sigma}+h.c.\right)
  +U\sum_i n_{i\uparrow}n_{i\downarrow},\label{H}
\end{eqnarray}
where $c_{i\sigma} (c_{i\sigma}^\dag)$ annihilates (creates) an electron
with spin $\sigma(=\uparrow, \downarrow)$ at the $i$th site and
$n_{i\sigma}=c_{i\sigma}^\dag c_{i\sigma}$.
$t$ denotes the nearest neighbor transfer integral 
and $U$ denotes the onsite Coulomb interaction.

\begin{figure}[htb]
 \centering
 \includegraphics[width=\linewidth]{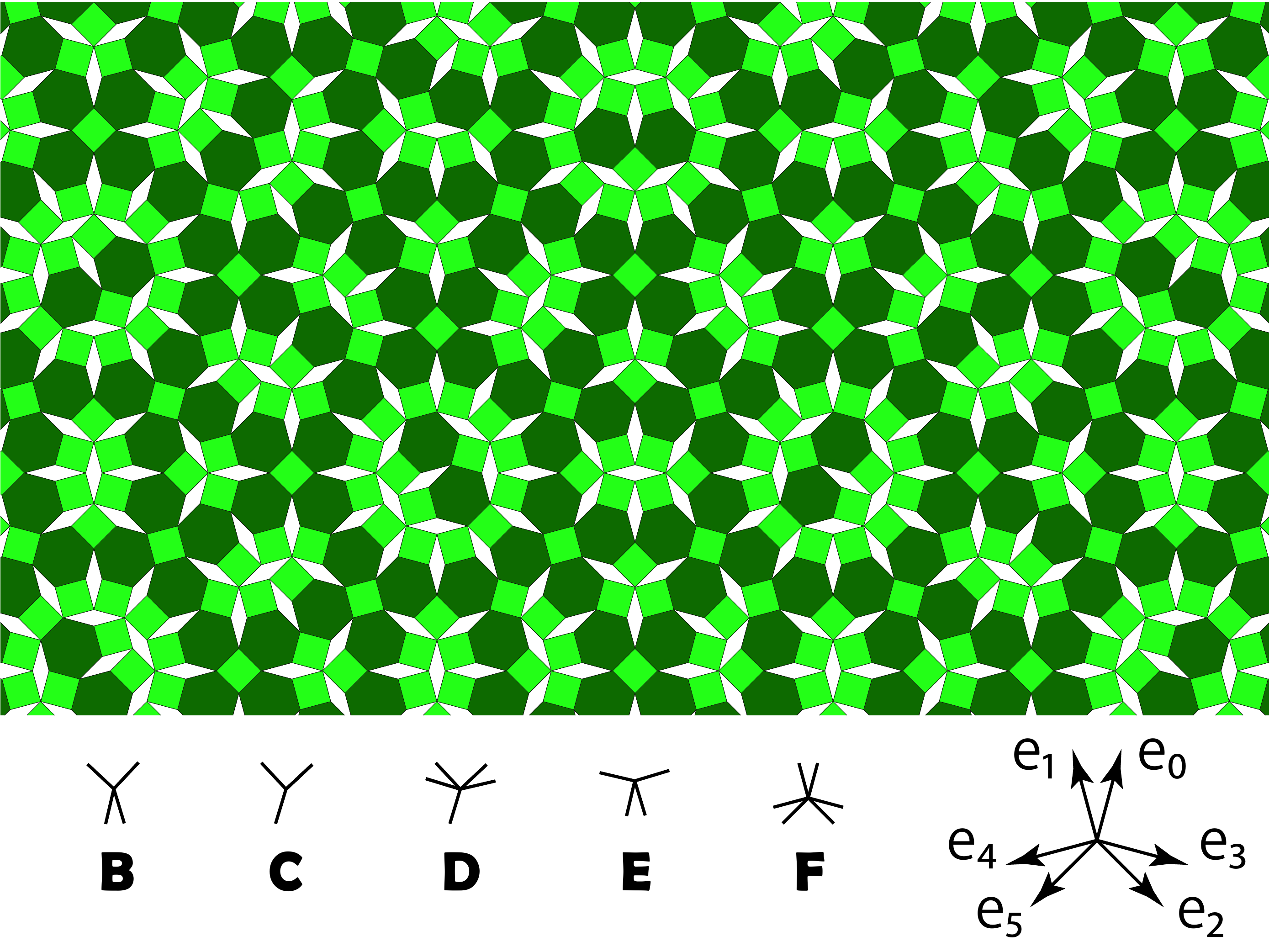}
 \caption{Socolar dodecagonal tiling and five types of vertices.
   ${\bf e}_0, {\bf e}_1, \cdots$, and ${\bf e}_5$ are
   projection of the fundamental
   translation vectors in six dimensions,
   ${\bf n}=(1,0,0,0,0,0), (0,1,0,0,0,0), \cdots$, and $(0,0,0,0,0,1)$.
 }
 \label{lattice}
\end{figure}

\begin{figure}[htb]
 \centering
 \includegraphics[width=\linewidth]{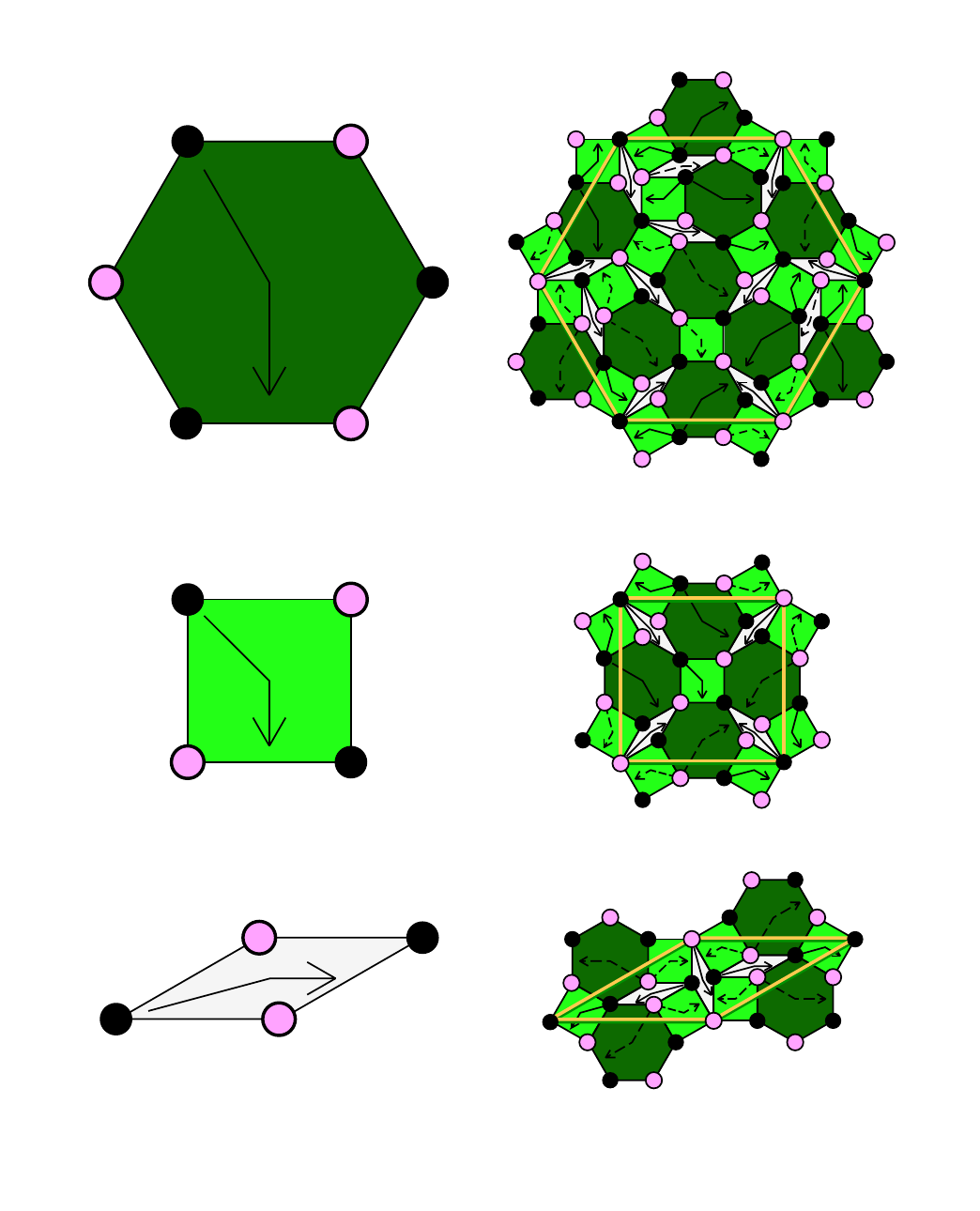}
 \caption{Deflation rule for the dodecagonal tiling~\cite{Socolar_1989}. 
 }
 \label{deflation}
\end{figure}

The Socolar dodecagonal tiling is
one of the two-dimensional quasiperiodic lattices
and is composed of hexagons, squares, and rhombuses,
which is schematically shown in Fig.~\ref{lattice}.
There are some methods to generate the quasiperiodic tilling
such as section and strip-projection methods.
In our study, we make use of the inflation-deflation rule
to generate the lattice~\cite{Socolar_1989}
since it is easy to obtain the exact fractions for various diagrams.
The deflation rules for directed hexagons, squares, and rhombuses,
are schematically shown in Fig.~\ref{deflation}.
It is known that the ratios of the hexagons, squares, and rhombuses are
$H:S:R=1:\sqrt{3}:\sqrt{3}$ in the thermodynamic limit~\cite{Socolar_1989}.
In the Socolar dodecagonal tiling, there are five kinds of vertices;
so-called B, C, D, E, and F vertices (see Fig.~\ref{lattice}).
We find that, in one deflation process,
B, C, D, and E vertices are newly generated in the inside of each hexagon, square, and rhombus,
as shown in Fig.~\ref{deflation}.
On the other hand, the corner sites of the hexagons, squares, and rhombuses
are changed into the F vertices under the deflation process.
Therefore, we obtain the fractions of the vertices in the thermodynamic limit as
\begin{eqnarray}
  p_B&=&\frac{3}{4}\left(5\sqrt{3}-9\right)\sim 0.255,\\
  p_C&=&\frac{3}{4}\left(\sqrt{3}-1\right)\sim 0.549,\\
  p_D&=&\frac{3}{4}\left(11\sqrt{3}-19\right)\sim 0.0394,\\
  p_E&=&\frac{1}{4}\left(9-5\sqrt{3}\right)\sim 0.0849,\\
  p_F&=&7-4\sqrt{3}=\tau^{-2}\sim 0.0718,
\end{eqnarray}
where $\tau(=2+\sqrt{3})$ is the ratio characteristic of the dodecagonal tiling.

Next, we consider the sublattice structure in the Socolar dodecagonal tiling
since we discuss the magnetically ordered state in the Hubbard model.
To this end, we introduce the sublattice-dependent vertices
B$_\sigma$, C$_\sigma$, $\cdots$, F$_\sigma$, where $\sigma$ is the sublattice index.
Its deflation rule is schematically shown in Fig.~\ref{deflation},
where the spin dependence of the vertices is represented
by the open and solid circles, and
the spin dependence of directed hexagons, squares, and rhombuses
are represented by the solid and dashed arrows.
Namely, we have defined the spin of one corner vertex,
which is located at the root of the arrow,
as the spin of the hexagons, squares, and rhombuses.
Then, the number of each hexagon, square, and rhombuses at iteration $n$
is changed under the deflation process,
which is explicitly given as ${\bf v}_{n+1}=M{\bf v}_n$,
with ${\bf v}_n^t=(H_\uparrow^{n}\,\, H_\downarrow^{n}\,\, S_\uparrow^{n}\,\, S_\downarrow^{n}\,\,
R_\uparrow^{n} \,\, R_\downarrow^{n})$ and
\begin{equation}
M=\left(
\begin{array}{cccccc}
  3 & 4 & 1+a & 1+a & 0  & 2-2a\\
  4 & 3 & 1+a & 1+a & 2-2a & 0\\
  6 & 6 & 1+3a & 3a & 2-3a & 4-3a\\
  6 & 6 & 3a & 1+3a & 4-3a & 2-3a\\
  6 & 6 & 2 & 2 & 2 & 1\\
  6 & 6 & 2 & 2 & 1 & 2
\end{array}
\right),
\end{equation}
where $H_\sigma^n$, $S_\sigma^n$, and $R_\sigma^n$ are
the number of hexagons, squares, and rhombuses with spin $\sigma$ at the iteration $n$,
and $a=2\sqrt{3}/9$.
The maximum eigenvalue of the matrix $M$ is $\tau^2$, and the corresponding eigenvector
is $(1, 1, \sqrt{3}, \sqrt{3}, \sqrt{3}, \sqrt{3})^t$.
Therefore, we can say that
no spin imbalance in the hexagons, squares, and rhombuses appears
in the thermodynamic limit.
As mentioned above, the B, C, D, and E vertices are generated
in the inside of the hexagons, squares, and rhombuses in each deflation operation
and thereby their fractions do not depend on the spin.
Furthermore, F vertices are changed from B, C, D, E, and F vertices
in each deflation operation.
This concludes no spin dependence in all vertices.
Therefore, we can say that
the antiferromagnetically ordered state is realized
without a net uniform magnetization in the thermodynamic limit~\cite{Lieb}.

\begin{figure}[htb]
 \centering
 \includegraphics[width=\linewidth]{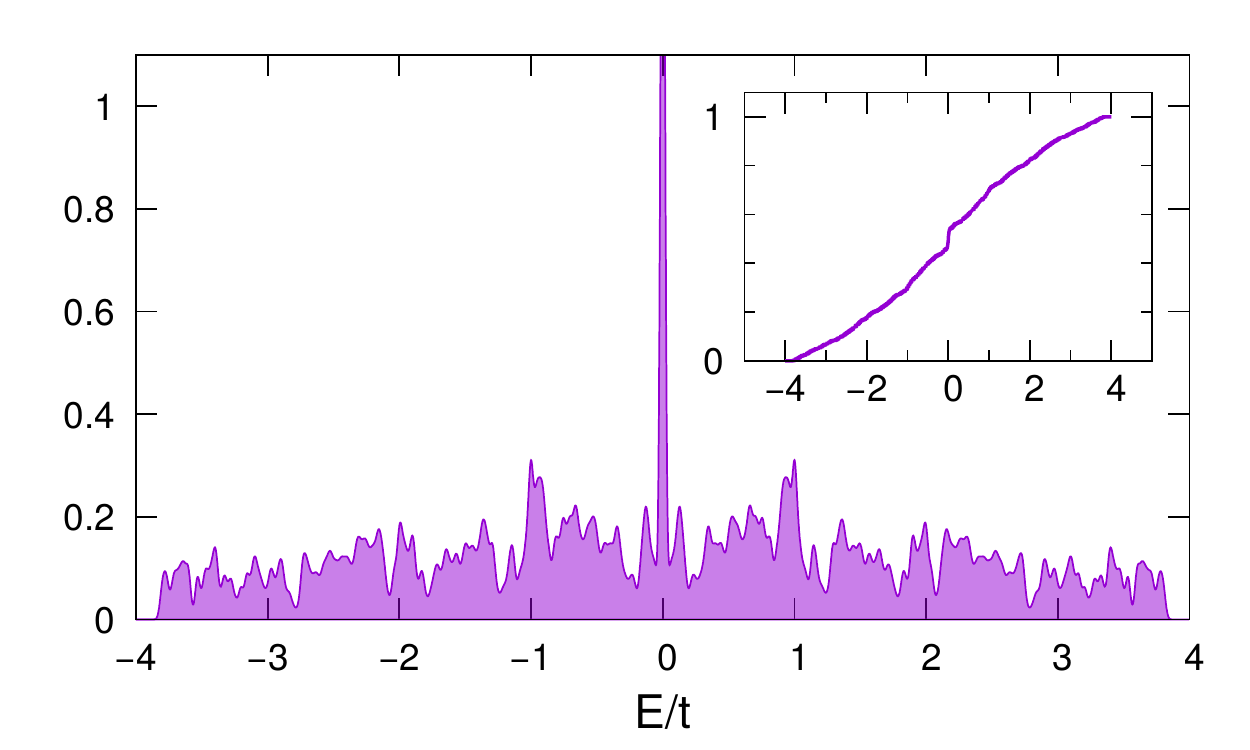}
 \caption{Density of states for the tightbinding model
   on the Socolar dodecagonal tiling with $N=290\,281$.
   }
 \label{dos}
\end{figure}

When the antiferromagnetically ordered state is considered in the quasiperiodic systems,
the density of states in the noninteracting case should play an important role
in the weak coupling regime.
In fact, the tightbinding model on the Penrose and Ammann-Beenker tilings has
a delta function like peak at $E=0$
and the introduction of Coulomb interactions drives the system
to the magnetically ordered states with finite magnetizations.
Now, we examine the density of states in the tightbinding model
on the Socolar dodecagonal tiling.
The results for the system with $N=290\,281$ are shown in Fig.~\ref{dos},
where $N$ is the number of site.
Since the vertex model on the Socolar dodecagonal tiling is bipartite,
the particle-hole symmetric DOS appears.
It is clearly found that the delta function like peak appears at $E=0$,
which suggests the existence of the confined states.
The fraction of the zero energy states is numerically obtained as $\sim 0.076$.
It is known that the value slightly depends on the system size
since there are zero energy states around the edge of the system
due to the open boundary condition.
Nevertheless, in the bulk, there indeed exist the confined states,
which is similar to the cases in the Penrose and Ammann-Beenker tilings.
In the next section, we examine the confined states
in the system in detail.

\section{Confined states}\label{conf}

In the section, we focus on the confined states with $E=0$
in the tightbinding model on the Socolar dodecagonal tiling.
Since a wave function described by the linear combination of
the confined states is also an eigenstate of the tightbinding Hamiltonian,
we can choose the simple state so that its occupied domain is small
and has a certain symmetry.
Then the domain is represented by a certain point group and
the confined state is described by its irreducible representation.
Since a certain domain repeats itself
in the quasiperiodic tiling,
the macroscopically degenerate states are realized
in the tightbinding model.
This is essentially the same as the Conway's theorem known
in the Penrose tiling.
Namely, the fractions of the confined states should be determined exactly.

\begin{center}
  \begin{table}
    \caption{A part of the irreducible characters of the point groups
      $D_3$, $D_2$, and $C_S$.
      $E$ is an identity operator, $C_3$ is a rotation operator of $2\pi/3$,
      and $I'$ is a reflection operator.}
    \begin{tabular}{crrr}
      \hline
      \hline
       $D_3$  & $E$ & $C_3$  & $I'$\\
      \hline
      A$_1$ &  1 &  1 & 1\\
      A$_2$ &  1 &  1 & -1\\
      \hline
    \end{tabular}
    \hspace{5mm}
    \begin{tabular}{crrrr}
      \hline
      \hline
       $D_2$  & $E$ & $C_2$  & $I_x$ & $I_y$ \\
      \hline
      A     &  1 &  1 & 1 & 1\\
      B$_1$ &  1 &  1 &-1 &-1\\
      B$_2$ &  1 & -1 & 1 &-1\\
      B$_3$ &  1 & -1 &-1 & 1\\
      \hline
    \end{tabular}
    \hspace{5mm}
    \begin{tabular}{crr}
      \hline
      \hline
       $C_S$  & $E$ & $I_y$\\
      \hline
      A$_1$ &  1 & 1\\
      A$_2$ &  1 & -1\\
      \hline
    \end{tabular}
    \label{D3}
  \end{table}
\end{center}

\begin{figure}[htb]
 \centering
 \includegraphics[width=\linewidth]{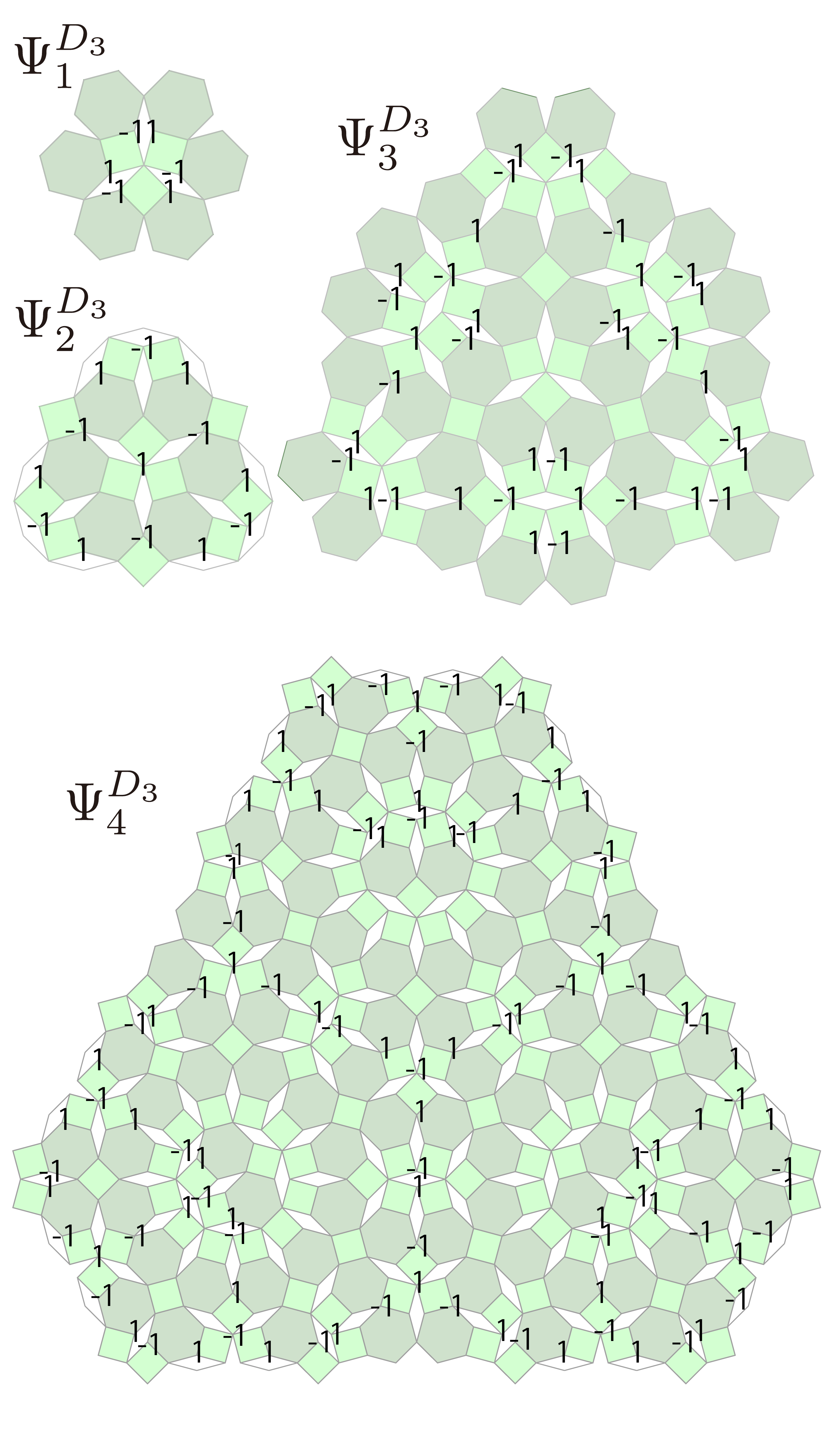}
 \caption{
   Four confined states in the domains with $D_3$ symmetry.
   The number at the vertices represent the amplitudes of the confined state.
 }
 \label{D3-2}
\end{figure}

\begin{figure}[htb]
 \centering
 \includegraphics[width=\linewidth]{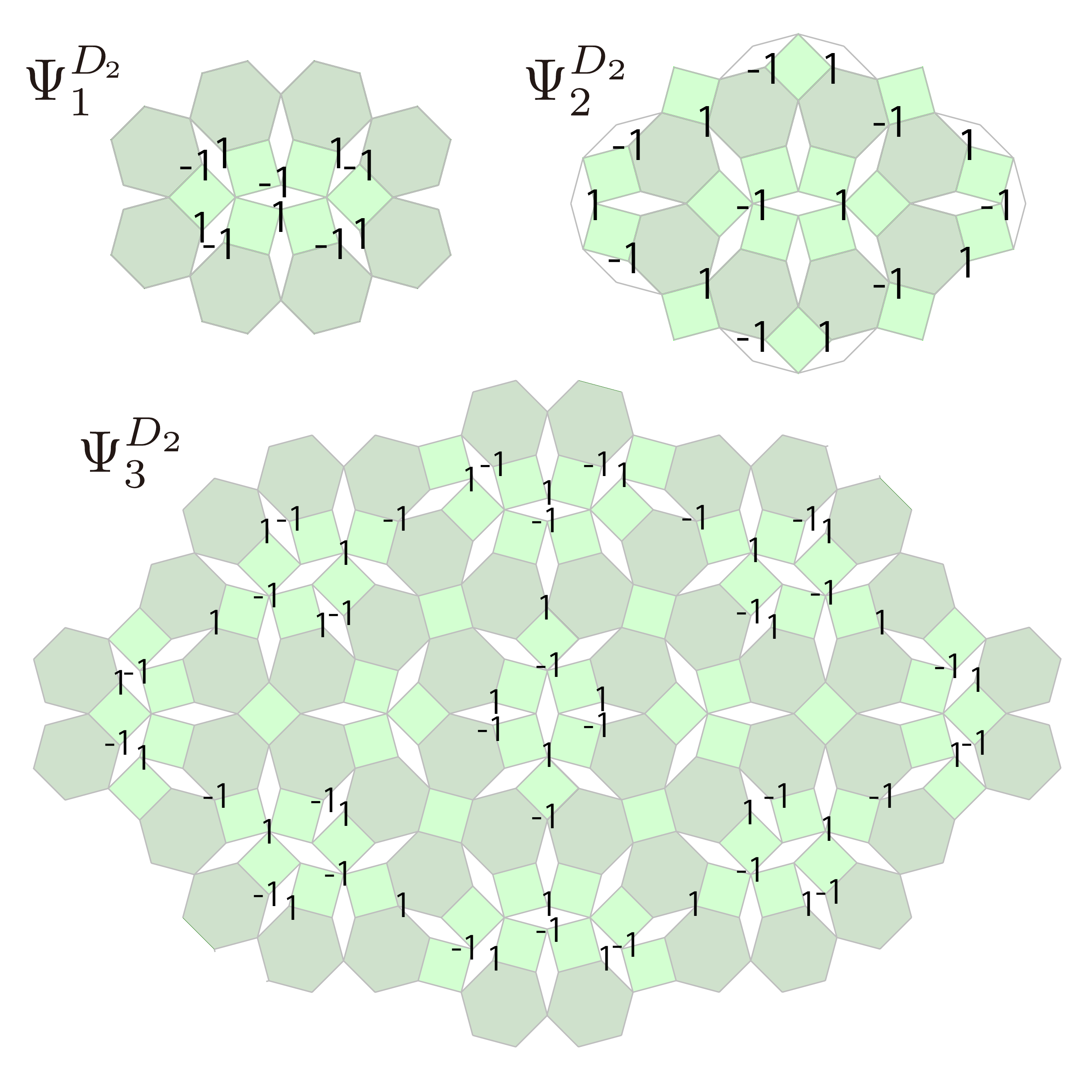}
 \caption{
   Three confined states in the domains with $D_2$ symmetry.   
   The number at the vertices represent the amplitudes of the confined state.
 }
 \label{D2-2}
\end{figure}

\begin{figure}[htb]
 \centering
 \includegraphics[width=\linewidth]{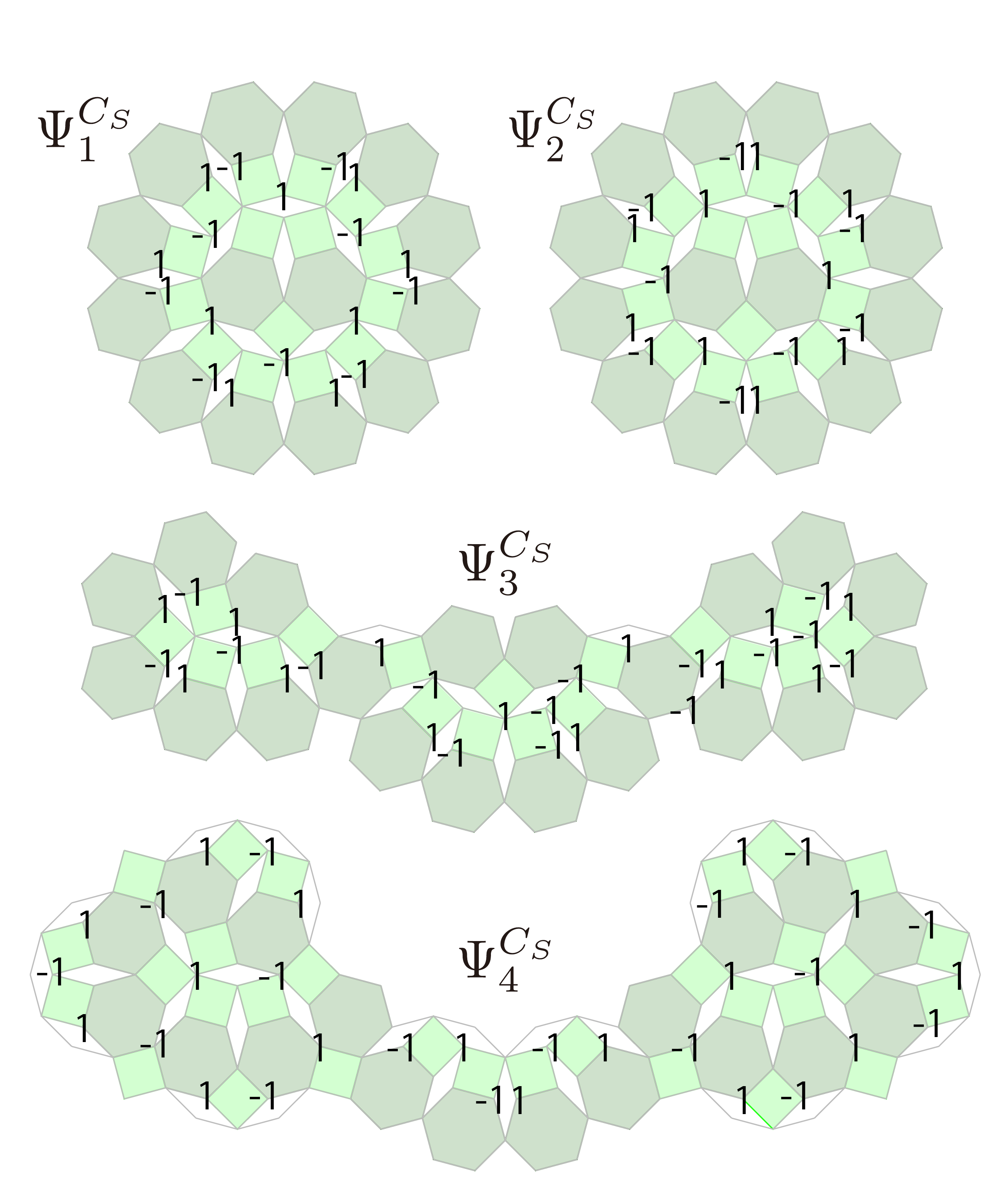}
 \caption{
   Four confined states in the domains with $C_S$ symmetry.   
   The number at the vertices represent the amplitudes of the confined state.
 }
 \label{CS-2}
\end{figure}

Now, we focus on the domain in the inside of
the Socolar dodecagonal tiling to examine the confined states (see Fig.~\ref{lattice}).
We find confined states in certain domains represented by
the $D_3$, $D_2$ or $C_S$ point group,
as shown in Figs.~\ref{D3-2}, \ref{D2-2}, and \ref{CS-2}, respectively.
It is also clarified that each confined states should be described by
the one-dimensional representation in the corresponding point group,
which is explicitly shown in Table~\ref{D3}.
The confined state $\Psi_1^{D_3}$ is described by
the representation A$_2$ in the $D_3$ point group, while
$\Psi_2^{D_3}$ is described by A$_1$, as shown in Fig.~\ref{D3-2}.
We have also found confined states in the larger domains
({\it eg.} $\Psi_3^{D_3}$ and $\Psi_4^{D_3}$),
which can not be described by the linear combinations of
the confined states in smaller domains.
This should suggest that, in the tightbinding Hamiltonian
on the Socolar dodecagonal tiling, there exist
infinite kinds of the confined states in larger domains.
This is similar to the Ammann-Beenker tiling,
while is different from the Penrose tiling.
The difference in these confined state properties
may be related to the following fact.
In the Penrose tiling, due to the existence of the ``cluster'' structure
bounded by ``forbidden ladders'',
each state is confined only in one cluster
and there are no confined states across the forbidden ladders.
This should yield a finite gap between the delta-function peak and
broad spectrum in the noninteracting density of states~\cite{Koga_Tsunetsugu_2017}.
On the other hand, in the Ammann-Beenker and Socolar dodecagonal tilings,
there are no ``forbidden regions'' and
therefore the multiple confined states have amplitudes in certain vertices.
This should yield energy levels close to the delta function peaks at $E=0$
since the corresponding state should have an amplitude in the whole system.
Namely, the gap size $\Delta$ is less than $0.002t$
in the system with $N=290\,281$
(see Fig.~\ref{dos}),
which is contrast to the Penrose tiling case with $\Delta\sim 0.17t$~\cite{Koga_Tsunetsugu_2017}.
This should lead to finite magnetization at almost all sites
in the Socolar dodecagonal tiling
even in the weak coupling limit.

\begin{figure}[htb]
 \centering
 \includegraphics[width=0.9\linewidth]{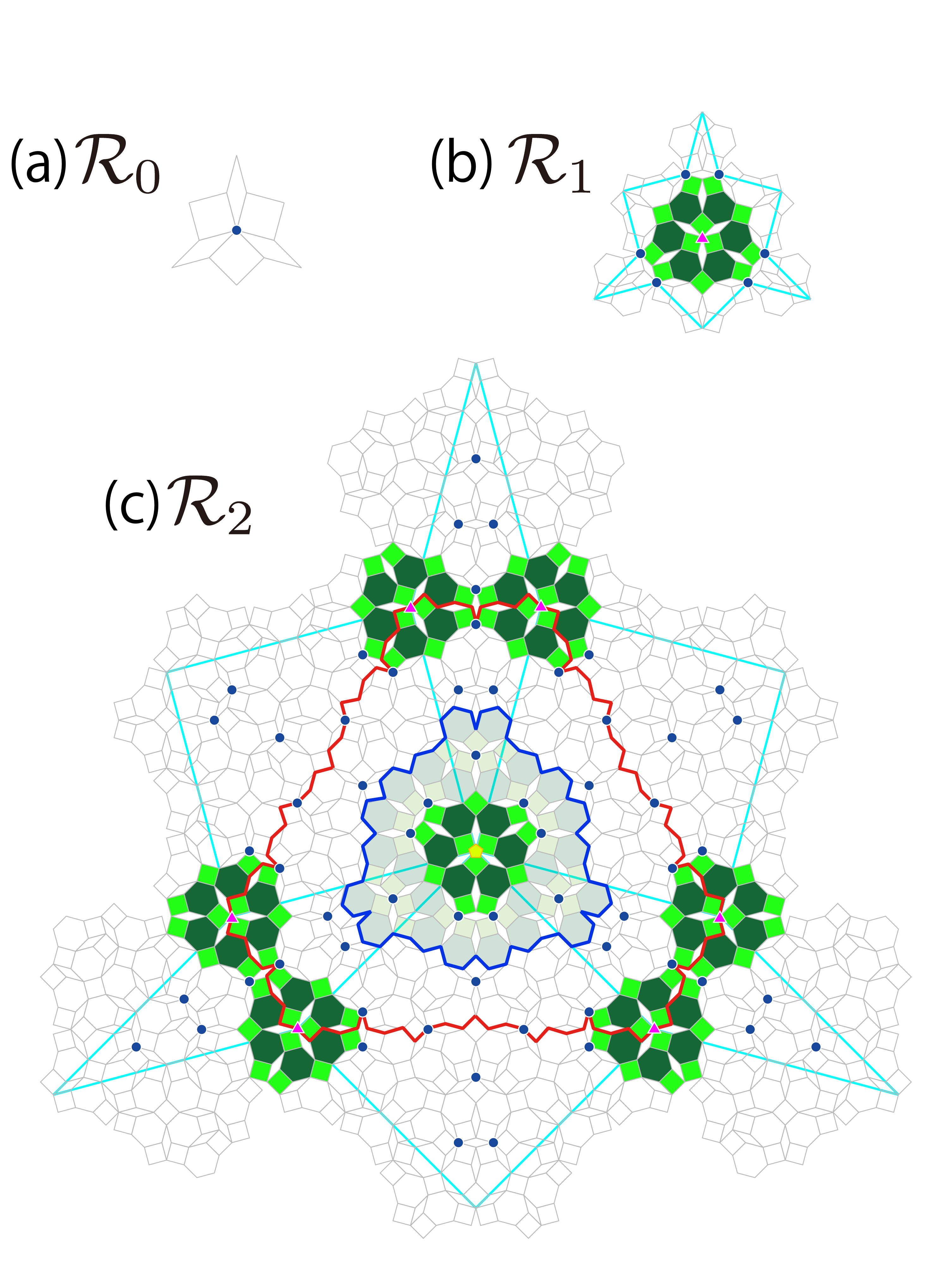}
 \caption{
   Regions ${\cal R}_0$, ${\cal R}_1$, and ${\cal R}_2$.
   Circles, triangles, and pentagon represent
   F$_0$, F$_1$, and F$_2$ vertices.
   Shaded regions (regions bounded by blue and red lines)
   represent the domains
   for $\Psi_1^{D_3}$ and $\Psi_2^{D_3}$ $\left(\Psi_3^{D_3}\right.$ and $\left.\Psi_4^{D_3}\right)$.
 }
 \label{F}
\end{figure}

Next, we examine the fraction of each confined state
by means of the inflation-deflation rule.
First, we consider the confined states described by the $D_3$ point group,
as shown in Fig.~\ref{D3-2}.
To this end, we consider the region composed of three squares
and three rhombuses (${\cal R}_0$),
which is invariant under $D_3$ symmetry operations,
as shown in Fig.~\ref{F}(a).
By applying the deflation process to the region ${\cal R}_0$, we obtain
the larger region ${\cal R}_1$, as shown in Fig.~\ref{F}(b).
We find that this region includes the structure of the region ${\cal R}_0$
around the center and is also invariant under the rotational symmetry.
Further application generates the region ${\cal R}_2$,
as shown in Fig.~\ref{F}(c).
These naturally expect that
applying the deflation process to the region ${\cal R}_i$,
the region ${\cal R}_{i+1}$,
in which the structure of ${\cal R}_i$ includes, is generated with $D_3$ symmetry.
Then, one can define the center site of the region ${\cal R}_i$ as an F$_i$ vertex,
where the integer $i$ strongly depends on the area with local $D_3$ symmetry.
Namely, in the ${\cal R}_i$ region, there should exist F$_j$ vertices
away from the center $(j<i)$,
which are shown as solid symbols in Fig.~\ref{F}.
The F$_0$ vertices are newly generated from the B, C, D, E vertices
in terms of the deflation operation, and 
the fraction of the F$_i$ vertices is
given as $p_{F_i}=(1-\tau^{-2})\tau^{-2i-2}$ in the thermodynamic limit.

Figure~\ref{F} shows that there exist the domains for the confined states
$\Psi_1^{D_3}$ and $\Psi_2^{D_3}$ around the F$_i$ vertices with $i\ge 1$,
and the domains for $\Psi_3^{D_3}$ and $\Psi_4^{D_3}$ appears around F$_i$ vertices
with $i\ge 2$.
Therefore, the corresponding fractions are obtained as,
\begin{eqnarray}
  p_1^{D_3}&=&p_2^{D_3}=\sum_{i=1}^\infty p_{F_i}=\frac{1}{\tau^4}\sim 0.0052,\\
  p_3^{D_3}&=&p_4^{D_3}=\sum_{i=2}^\infty p_{F_i}=\frac{1}{\tau^6}\sim 0.0037.
\end{eqnarray}

As for the confined states specified by $D_2$ and $C_S$ point symmetries,
it may be convenient to consider the deflation rule for the hexagons, squares, and
rhombuses.
Namely, no confined states specified by $D_2$ or $C_S$ symmetry are obtained
by one deflation process (see Fig.~\ref{deflation}).
\begin{figure}[htb]
 \centering
 \includegraphics[width=0.9\linewidth]{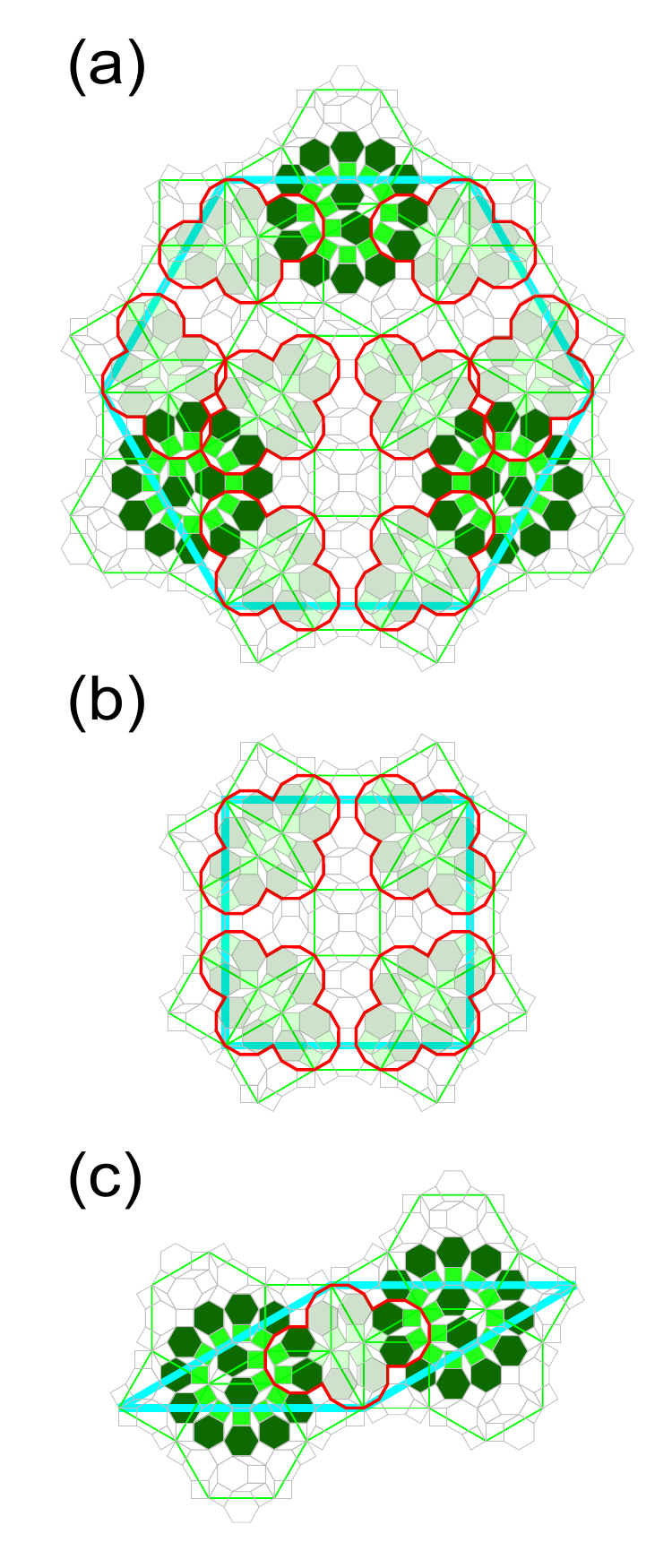}
 \caption{
   Two deflation processes for (a) hexagon, (b) square, and (c) rhombus.
   The domains for the confined states $\Psi_1^{D_2}$ and $\Psi_1^{C_S}$
   are shown as shaded regions.
 }
 \label{def2}
\end{figure}
Figure~\ref{def2} shows clusters obtained from two deflation processes to
one hexagon, square, and rhombus.
We find some smallest domains
for the confined states $\Psi_1^{D_2}$ and $\Psi_2^{D_2}$
($\Psi_1^{C_S}$ and $\Psi_2^{C_S}$).
Therefore, the fractions for the confined states are obtained as
\begin{eqnarray}
  p_{1}^{D_2}=p_2^{D_2}&=&\frac{1}{\tau^4}\left(8p_H+4p_S+p_R\right)
  =\frac{1}{\tau^3}\sim 0.019,\\
  p_{1}^{C_S}=p_2^{C_S}&=&\frac{1}{\tau^4}\left(3p_H+2p_R\right)
  =\frac{\tau^2+1}{11\tau^4}\sim 0.0070,
\end{eqnarray}
where the fractions of hexagons, squares, and rhombuses are
given as $p_H=1/(2\sqrt{3}+1)$ and $p_S=p_R=\sqrt{3}/(2\sqrt{3}+1)$.
The confined states in larger domains, which are shown in Figs.~\ref{D2-2} and \ref{CS-2},
should appears in the larger regions generated
by multiple deflation processes.

In the study, we could find some confined states in smaller domains.
Therefore, we can give the lower bound of the fraction of the confined states as,
\begin{eqnarray}
  p>p_{lower}&=&\sum_{i=1}^4p_i^{D_3}+\sum_{i=1}^2p_i^{D_2}+\sum_{i=1}^2p_i^{C_S}\\
  &=&\frac{2}{11\tau^5}\left(60\tau+29\right)\sim 0.064.
\end{eqnarray}
It is found that the obtained lower bound is consistent
with the fraction $0.076$ in the system with $N=290281$.
In the following, we clarify how the introduction of the Coulomb interactions
realizes the antiferromagnetically ordered states.

\section{Effect of the Coulomb interaction}\label{results}
In the section, we consider the Hubbard model with finite $U$.
To study the antiferromagnetically ordered state characteristic of
the Socolar dodecagonal tiling,
we use the real-space Hartree approximation and
the Hamiltonian (\ref{H}) is reduced to
\begin{eqnarray}
  H_{MF}&=&-t\sum_{\langle ij\rangle \sigma}\left(c_{i\sigma}^\dag c_{j\sigma}+h.c.\right)
  +U\sum_{i\sigma}\langle n_{i\bar{\sigma}}\rangle n_{i\sigma},
\end{eqnarray}
where $\langle n_{i\sigma} \rangle$ is the expectation value of
the number of electron with spin $\sigma$ at the $i$th site.
In our calculations, we deal with finite lattices with $N=72\,274$ and $290\,281$
under the open boundary condition.
The lattices are generated by applying the deflation processes
to the F vertex.
Therefore, the obtained vertex lattices have
the global threefold rotational symmetry.
For given values of mean-fields, we numerically diagonalize
the mean-field Hamiltonian
$H_{MF}$ and update the mean-fields, and iterate this selfconsistent procedure
until the result converges within numerical accuracy.

We show in Fig.~\ref{magpat} the spatial pattern of the magnetization
$m_i(=\langle n_{i\uparrow}\rangle - \langle n_{i\downarrow}\rangle)/2$
when $U/t=1.0\times 10^{-7}$, 1, 2, and 5.
\begin{widetext}
  \begin{center}
    \begin{figure}[htb]
      \includegraphics[width=\textwidth]{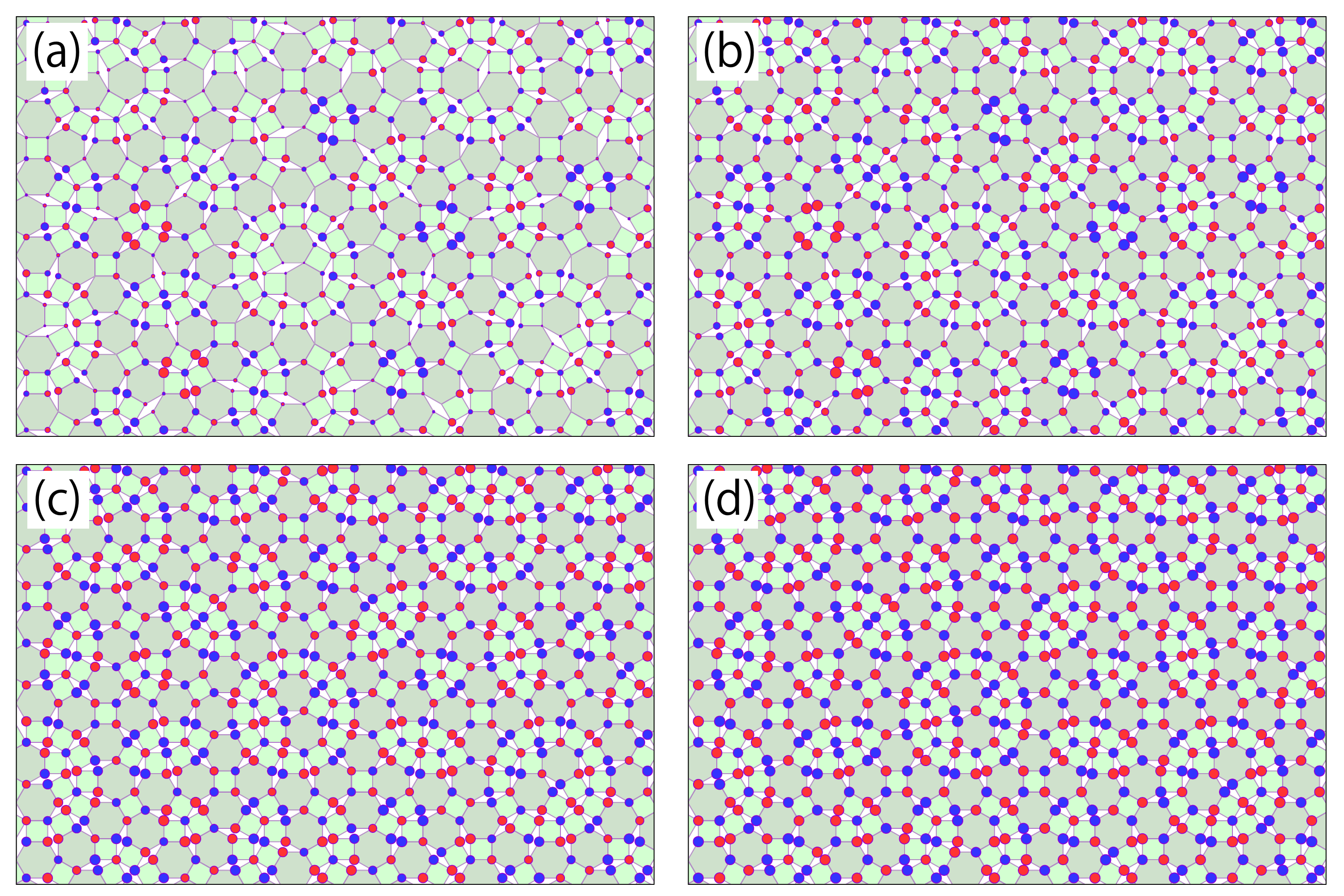}
      \caption{Spatial pattern for the staggered magnetization
        in the Hubbard model on the Socolar dodecagonal tiling
        when $U/t=1.0\time 10^{-7}$ (essentially the same as $U=0$),
        1, 2, and 5. The area of the circles represents
      the normalized magnitude of the local magnetization.}
      \label{magpat}
    \end{figure}
  \end{center}
\end{widetext}
It is found that finite staggered magnetizations are induced even in the weak coupling limit
$U\rightarrow 0$.
This originates from the existence of the confined states at $E=0$,
as discussed above.
In fact, the larger magnetizations appear at the vertices,
where the confined states $\Psi_1^{D_2}$ and $\Psi_1^{D_3}$ have amplitudes.
Increasing the Coulomb interactions,
local magnetizations monotonically increase.
When $U/t=5$, the staggered magnetizations are almost polarized.
It is known that, in the strong coupling case, the system is reduced to the Heisenberg model
on the Socolar dodecagonal tiling with nearest neighbor exchanges $J=4t^2/U$.
The ground state obtained by the mean-field theory is fully polarized
with the staggered moment $m_j=\pm 1/2$ in the limit.
This is different from the results obtained by the spin wave theory for the Heisenberg model,
which predicts site-dependent reduction in the spontaneous moments.
This reduction originates from intersite quantum fluctuations.
Although intersite correlations could not be taken into account
in the framework of the mean-field approximations,
the essence of crossover behavior in the magnetic properties
can be captured correctly.

Finally, we consider the magnetization profile in the perpendicular space.
The positions in the perpendicular space have one-to-one correspondence
with the positions in the physical space.
The vertex sites in the Socolar dodecagonal tiling correspond to
a subset of the six-dimensional lattice points $\vec{n}=(n_0, n_1, n_2, n_3, n_4, n_5)$
labeled with integers $n_m$ and their coordinates are the projections onto
the two-dimensional physical space:
\begin{eqnarray}
  {\bf r}=(x,y)=\sum_{m=0}^5 n_m {\bf e}_m,
\end{eqnarray}
where ${\bf e}_m=(\cos (m\theta+\theta_0), \sin (m\theta+\theta_0))$ for $m=0, 1, 4, 5$,
${\bf e}_m=-(\cos (m\theta+\theta_0), \sin (m\theta+\theta_0))$ for $m=2, 3$,
$\theta=2\pi/3$ and
the initial phase $\theta_0$ is arbitrary,
and an example with the choice $\theta_0=5\pi/12$.
The projection onto the four-dimensional perpendicular space has information
specifying the local environment of each site~\cite{Socolar_1989},
\begin{eqnarray}
  \tilde{\bf r}&=&\sum_{m=0}^5n_m\tilde{\bf e}_m,\\
  {\bf r}^\perp&=&\sum_{m=0}^5n_m{\bf e}^\perp_m,
\end{eqnarray}
where $\tilde{\bf e}_m={\bf e}_{5m ({\rm mod\, 12})}$,
${\bf e}^\perp_m=(\delta_{m (\rm mod\, 2),0},\delta_{m (\rm mod\,2),1})$.
${\bf e}^\perp$ takes only four values $(0,0), (0,1), (1,0), (1,1)$, and
in each ${\bf e}^\perp$ plane the $\tilde{\bf r}$ points densely cover a region
with hexagonal shape.
The areas and their structures do not depend on planes,
which is contrast to the perpendicular space of the Penrose tiling
with two distinct planes.
Note that the sites with odd and even number of $l={\bf e}^\perp \cdot (1,1)$ correspond
to the A/B sublattices since upon moving from one site to its neighbor site
only one of $n_m$'s changes by $\pm 1$.
Therefore, the antiferromagnetically ordered state may also be characterized by
an alternate sign of the magnetization in the four planes.
Then, we can discuss the magnetization profile only in one plane,
taking into account the equivalent planes and sign of the magnetization. 
Figure~\ref{perp}(d) shows one of the planes in the perpendicular space,
where each part is the region of the five kinds of vertices.
We also find that the F vertices are densely distributed
in the hexagonal structure at the center of the perpendicular space.
This nested (fractal) structure means that
the set of F vertices is the Socolar dodecagonal tiling
with larger lattice constant.
This is one of the important features characteristic of the Socolar dodecagonal tiling.

The magnetization profile is shown in Fig.~\ref{perp}.
In the weak coupling limit, the macroscopically degenerate confined states
yield the spatial distribution in the magnetization,
which leads to interesting magnetization profile even in the perpendicular space.
On the other hand, we could not find the nested structure in the magnetization profile.
This is contrast to the magnetic profile in the Hubbard model on
the Ammann-Beenker tiling~\cite{ABKoga}.
This may originate from the following.
In the Ammann-Beenker tiling case,
the confined state located in smaller domains has no amplitudes
in the locally symmetric vertices (F vertices),
while the confined state has a finite amplitude at the vertices
away from the center
in the larger domain.
The confined states in larger domains have a tiny amplitude in F vertices
and their fractions are small,
which should lead to the superlattice structure in the magnetic profile.
On the other hand, in the Socolar dodecagonal tiling,
the confined state $\Psi_2^{D_3}$ in the smaller domain
has an amplitude at the center F vertex,
as shown in Fig.~\ref{D3-2}.
Therefore, relatively larger magnetization $|m|\sim 1/26$ is induced
although its magnitude should be slightly modified by taking into account
the hybridization to other confined states at larger domains.

Increasing $U$, the interesting magnetic pattern smears and
magnetic properties become classified by vertices, as shown in Fig.~\ref{perp}.
In fact, it seems that, in the perpendicular space,
the region of the five kinds of vertices has almost a single staggered magnetization
in the system with $U/t=5$, as shown in Fig.~\ref{perp}(c).
Then we could find the crossover behavior in the magnetic profile.
\begin{widetext}
  \begin{center}
    \begin{figure}[htb]
      \includegraphics[width=\textwidth]{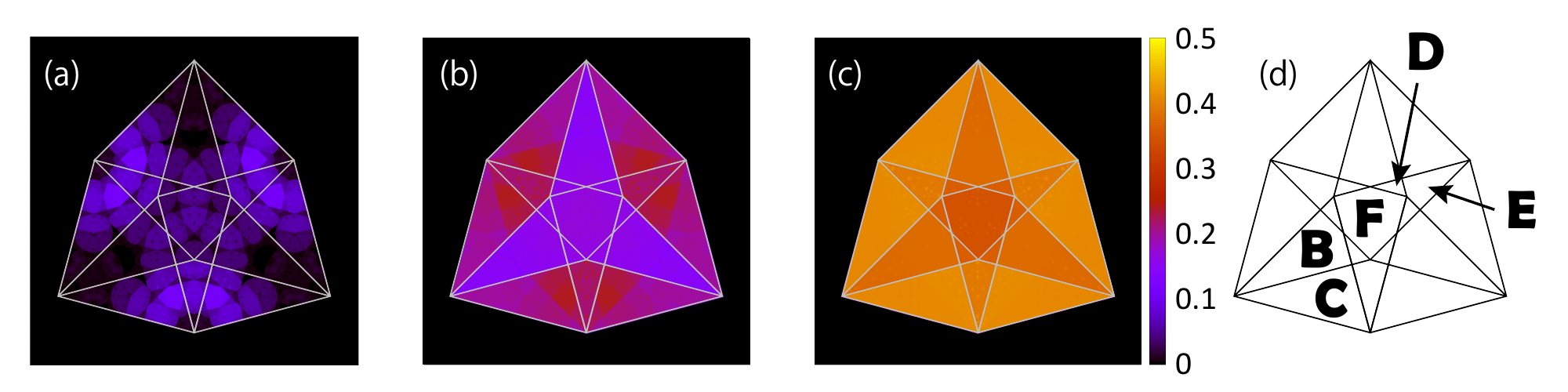}
      \caption{Magnetization profile in the perpendicular space
        for the system with $N=290\,281$
        when $U/t= (a) 1.0\time 10^{-7}$ (essentially the same as $U=0$),
        (b) 2, and (c) 5.
        (d) Each part is the region of the five kinds of vertices
        shown in Fig.~\ref{lattice}.
        }
      \label{perp}
    \end{figure}
  \end{center}
\end{widetext}

\section{Summary}
We have studied the antiferromagnetically ordered state
in the half-filled Hubbard model
on the Socolar dodecagonal tiling.
When the interaction is introduced,
the staggered magnetizations suddenly appear,
which results from the existence of the macroscopically degenerate states in
the tightbinding model.
By examining the confined states with $E=0$ in detail,
we have determined the lower bound of their fraction.
The increase of the interaction strength monotonically increases
the magnetizations.
We have also mapped the spatial distribution of the magnetization
to the perpendicular space, and have discussed the magnetization profile.
It has been clarified that
in the strong coupling regime,
the staggered magnetizations are almost classified into five vertices,
depending on the site coordination numbers.
On the other hand, in the weak coupling regime,
the interesting pattern appears in the perpendicular space,
which originates from the existence of the confined states.

It has been clarified that the magnetic profile characteristic of the Socolar dodecagonal tiling clearly appears in the real space. Therefore, it may be difficult to observe in the neutron inelastic scattering experiments (the Fourier space). However, we expect that the beautiful magnetic pattern will be observed in the detailed analysis in the spin-dependent scanning tunneling microscopy measurements for the magnetic quasicrystals.

\begin{acknowledgments}
  We would like to thank S. Sakai for valuable discussions.
  Parts of the numerical calculations are performed
  in the supercomputing systems in ISSP, the University of Tokyo.
  This work was supported by Grant-in-Aid for Scientific Research from
  JSPS, KAKENHI Grant Nos.
  JP19H05821, JP18K04678, JP17K05536 (A.K.).
\end{acknowledgments}

\bibliography{./refs}

\end{document}